# Generalized hyper-Ramsey resonance with spinors

Thomas Zanon-Willette, Alexey V. Taichenachev, Valeriy I. Yudin

*Abstract.* The generalized hyper-Ramsey resonance formula originally published in Phys. Rev. A vol 92, 023416 (2015) is derived using a Cayley-Klein spinor parametrization. The shape of the interferometric resonance and the associated composite phase-shift are reformulated including all individual laser pulse parameters. Potential robustness of signal contrast and phase-shift of the wave-function fringe pattern can now be arbitrarily explored tracking any shape distortion due to systematic effects from the probe laser. An exact and simple analytical expression describing a Ramsey's method of separated composite oscillating laser fields with quantum state control allows us to accurately simulate all recent clock interrogation protocols under various pulse defects.

## 1. Introduction

In 1949, N.F. Ramsey developed a separated oscillating fields method in magnetic resonance in order to upgrade the Rabi pulse technique measuring atomic transition frequencies with higher precision [1,2]. The Ramsey's method was thus intensively applied in time and frequency metrology leading to modern atomic clocks with the highest precision [3]. The method was also extended to matter-wave interferometry using the atomic recoil effect induced by lasers to realize beam splitters and mirrors [4]. Atomic gravito-inertial sensors based on Ramsey-Bordé interferometers have thus rapidly improved measurement sensitivity to external fields or rotations [5].

This short communication focuses on quantum engineering of a Doppler-recoil-free laser pulsed Ramsey resonance associated with the atomic phase-shift driven by a sequence of three composite pulses applied around a single free evolution time [6,7]. The original scheme called hyper-Ramsey (HR) spectroscopy demonstrates a very strong efficiency to eliminate the sensitivity to residual probe induced light-shift in a single $^{171}$Yb$^+$ ion clock [8]. A new frequency standard based on the very narrow electric octupole (E3) transition leads to an unprecedented relative accuracy at $3 \times 10^{-18}$ [9].

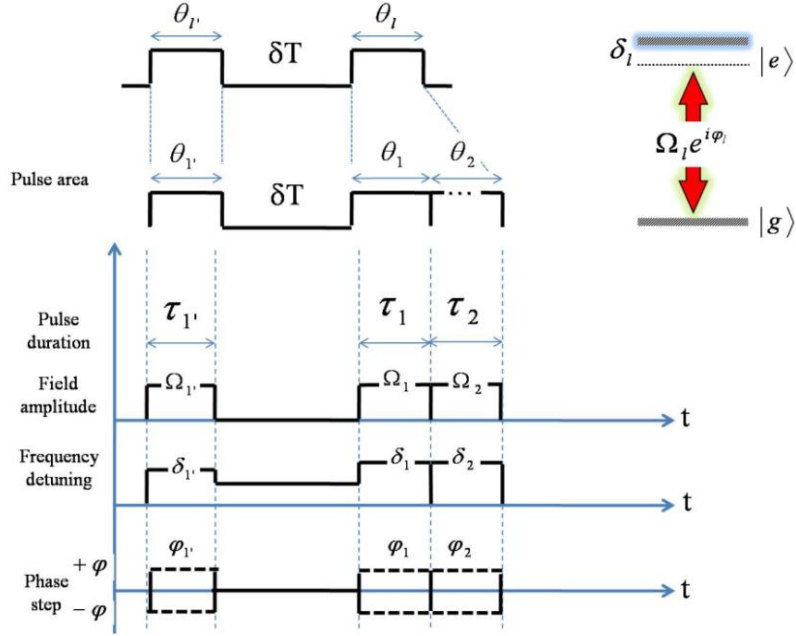

**Figure 1. Composite three-pulse spectroscopy probing an ultra-narrow clock transition.** Optical pulses are defined by a generalized area $\widetilde{\theta}_l$ ($l = 1'; l = 1, 2$) the frequency detuning $\delta_l$, the Rabi field frequency $\Omega_l e^{i\varphi_l}$ including a laser phase-step, a pulse duration $\tau_l$ and a single free evolution time T. The general clock frequency detuning is defined by $\delta_l = \delta - \Delta_l$ where $\Delta_l$ is the residual error after pre-compensation of the laser probe induced frequency shift.

Such a beautiful result has thus stimulated intensive research in the field of Ramsey spectroscopy leading to other sequences of composite pulses even more robust to residual light- shift [10]. An exact analytical transition probability denoted as the generalized hyper-Ramsey resonance was initially presented in [7]. Such a resonance was an extension of the original hyper-Ramsey scheme taking into account a few modifications of laser parameters during laser probe interrogation. But the exact influence of the laser phase-step on each pulse was not clearly provided and the effect of a residual coherence was missing.

We have reported in this paper, for the first time, a new derivation of the same resonance by the spinor formalism. An exact and simple closed form solution for both the resonance shape and the central fringe phase-shift are established including all laser parameters from each individual tailored pulse. The spinor matrices formalism would provide an accurate quantum simulation platform testing other sequences of composite pulses within Ramsey spectroscopy dedicated to a next generation of robust atomic clocks and spectrometers against laser probe perturbations [10,11].

We start the formal derivation of the generalized hyper-Ramsey (GHR) transition probability by introducing Cayley-Klein parametrization of rotation spinors as following [10,12,13]:

$$M(\widetilde{\vartheta}_l) = \begin{pmatrix} \cos\widetilde{\vartheta}_l\, e^{i\phi_l} & -ie^{-i\varphi_l}\sin\widetilde{\vartheta}_l \\ -ie^{i\varphi_l}\sin\widetilde{\vartheta}_l & \cos\widetilde{\vartheta}_l\, e^{-i\phi_l} \end{pmatrix} \quad (1)$$

where we use angle reduced notation $\tilde{\theta}_l = \theta_l/2$ and $\tilde{\vartheta}_l = \vartheta_l/2$ and introducing the effective pulse area as $\theta_l = \sqrt{\delta_l^2 + \Omega_l^2}\,\tau_l$. Phase angles are introduced by the following definitions:

$$\vartheta_l = 2\arcsin\left[\frac{\Omega_l}{\omega_l}\sin\tilde{\theta}_l\right] \quad (2a)$$

$$\phi_l = \arctan\left[\frac{\delta_l}{\omega_l}\tan\tilde{\theta}_l\right] \quad (2b)$$

Such a parametrization is very convenient to simulate any composite three-pulse configuration shown in Fig.1. Applying spinors to the Schrödinger equation with $c_g(0), c_e(0)$ as initial conditions, a closed form solution is derived for the complex GHR amplitude as following:

$$\begin{pmatrix}c_g\\c_e\end{pmatrix} = M(\tilde{\vartheta}_2)\cdot M(\tilde{\vartheta}_1)\cdot\begin{pmatrix}e^{i\frac{\delta T}{2}} & 0\\ 0 & e^{-i\frac{\delta T}{2}}\end{pmatrix}\cdot M(\tilde{\vartheta}_{1'})\cdot\begin{pmatrix}c_g(0)\\c_e(0)\end{pmatrix} \quad (3)$$

The clock transition can be detected by measuring the atomic population fraction in the ground state $P_g = |c_g|^2$ or in the excited state $P_e = |c_e|^2 = 1 - P_g$.

## 2. Exact generalized hyper-Ramsey (GHR) resonance formula

The corresponding transition probability of the ground state for initial population initialization given by $c_g(0) = 1, c_e(0) = 0$ can be recast in the canonical form as:

$$P_g = |\alpha|^2\left|\left(1 - |\beta|e^{-i(\delta T + \Phi)}\right)\right|^2 \quad (4a)$$

including a composite phase-shift accumulated over the entire sequence of laser pulses given by:

$$\Phi = \varphi_1 - \varphi_{1'} + \phi. \quad (4b)$$

with an additional complex phase-shift contribution:

$$\phi = \phi_{1'} + \phi_1 - \arg[\beta] \quad (4c)$$

Parameters $\alpha$ and $\beta$ driving the overall envelop and composite phase-shift can be separated in two contributions from left pulse driven by pulse area $\tilde{\vartheta}_{1'}$ and right pulses driven by $\tilde{\vartheta}_1, \tilde{\vartheta}_2$ around the free evolution time matrix. We introduced then the reduced notations:

$$\alpha \equiv \alpha(\tilde{\vartheta}_{1'})\,\alpha(\tilde{\vartheta}_{12}) \quad (5a)$$

$$\beta \equiv \beta(\tilde{\vartheta}_{1'})\,\beta(\tilde{\vartheta}_{12}) \quad (5b)$$

where

$$\alpha(\tilde{\vartheta}_{1'}) = \cos\tilde{\vartheta}_{1'} \qquad (6a)$$

$$\alpha(\tilde{\vartheta}_{12}) = \cos\tilde{\vartheta}_1 \cos\tilde{\vartheta}_2 \left(1 - e^{-i\Xi_{12}} \tan\tilde{\vartheta}_1 \tan\tilde{\vartheta}_2\right) \qquad (6b)$$

$$\beta(\tilde{\vartheta}_{1'}) = \tan\tilde{\vartheta}_{1'} \qquad (6c)$$

$$\beta(\tilde{\vartheta}_{12}) = \frac{\tan\tilde{\vartheta}_1 + e^{-i\Xi_{12}}\tan\tilde{\vartheta}_2}{1 - e^{-i\Xi_{12}}\tan\tilde{\vartheta}_1\tan\tilde{\vartheta}_2} \qquad (6d)$$

with $\Xi_{12} = \varphi_2 - \varphi_1 + \phi_1 + \phi_2$.

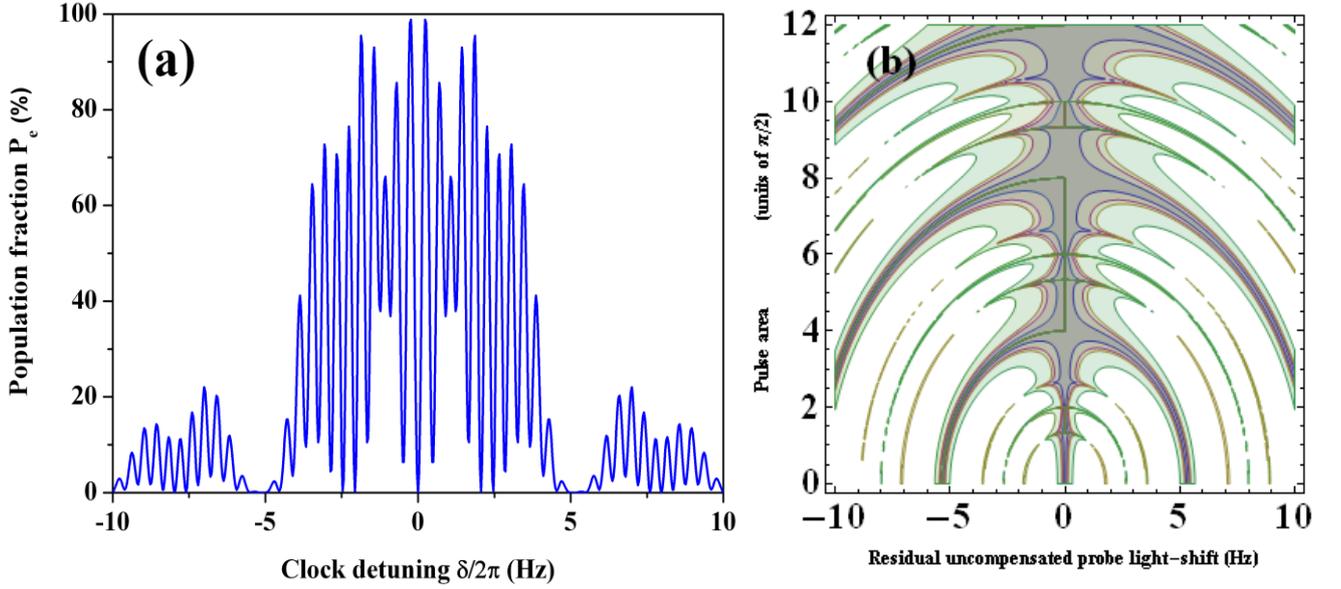

**Figure 2.** Population fraction $P_e$ and associated clock frequency shift versus the clock detuning. (a) Hyper-Ramsey (HR) resonance. (b) 2D map representation of the HR clock frequency-shift $\delta\nu$ plotted versus residual uncompensated probe light-shift $\Delta/2\pi$ (horizontal axis) and versus the pulse area in units of $\pi/2$ (vertical axis) for initial conditions $c_g(0) = 1, c_e(0) = 0$. Colored contour plots are drawn by superposition of layers for different frequency shifts starting from 10 mHz (light zone) to below 0.05 mHz (dark zone). A superposition of layers leading to a dark region means a very low sensitivity < 0.05 mHz to residual probe light-shift. Spectroscopic parameters are the Rabi field $\Omega = \pi/2\tau$ with a pulse duration $\tau = 3/16$ s, a free evolution time T=2s and a laser probe-induced residual frequency shift as $\Delta_{1'} \equiv \Delta_1 \equiv \Delta_2 = \Delta$.

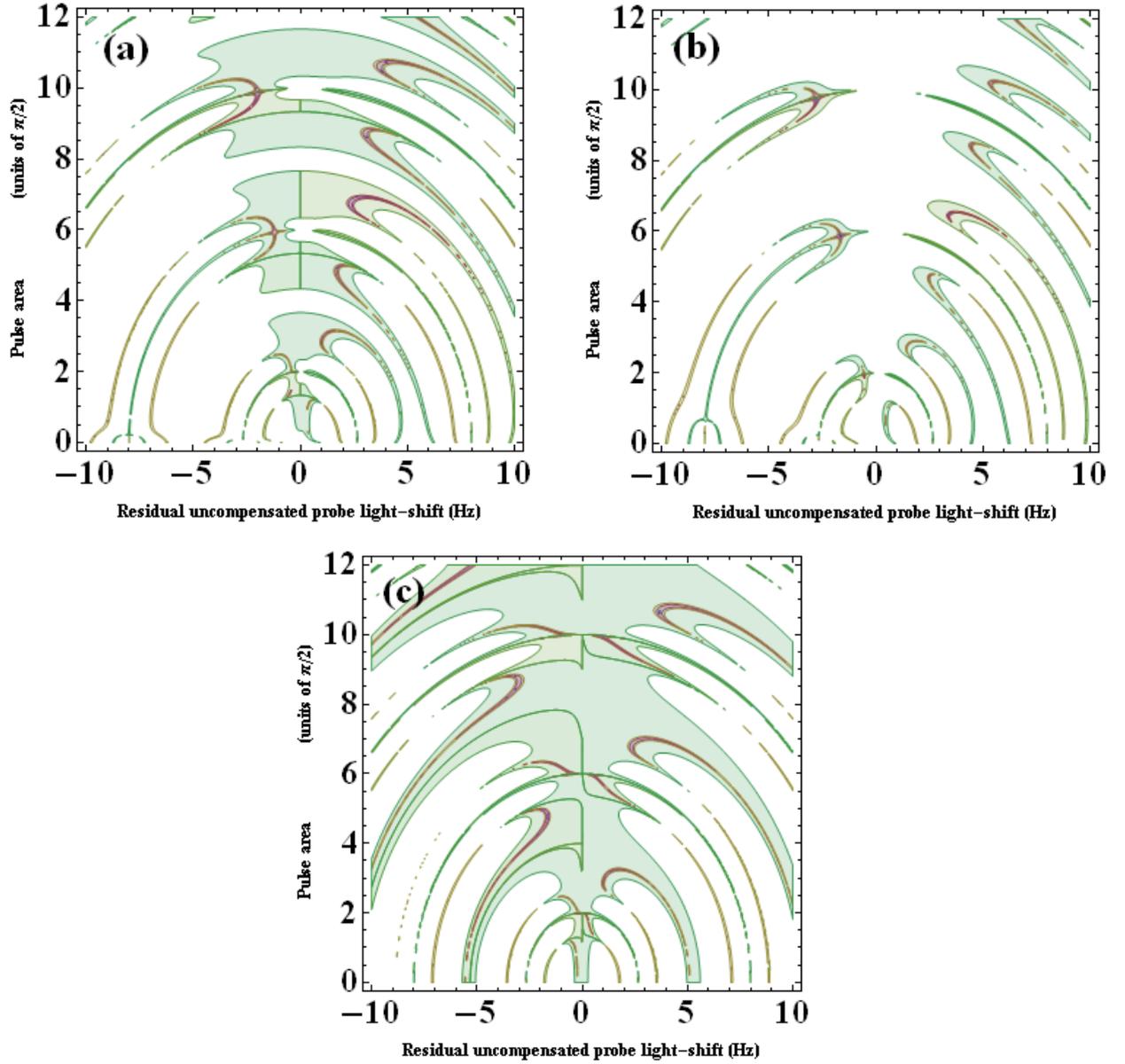

**Figure 3.** Distortion of the 2D map representation of the HR clock frequency-shift δν from $P_e$ due to imperfect state initialization or phase-step error. The frequency-shift related to Eq. 12 is plotted versus residual uncompensated probe light-shift $\Delta/2\pi$ (horizontal axis) and versus the pulse area in units of $\pi/2$ ( (vertical axis). (a) $c_g(0) = \sqrt{0.999}, c_e(0) = \sqrt{0.001}$, (b) $c_g(0) = \sqrt{0.99}, c_e(0) = \sqrt{0.01}$, (c) Phase-step error $\varphi_1 = 0.99\,\pi$ during the middle pulse. Others parameters and colored contour plots as in Fig. 2.

From Eq. 4c, few composite phase-shift formulae can be straightforwardly recovered as following:

Identical two pulse Ramsey phase-shift:

$$\phi = 2\phi_{1'} \qquad (8a)$$

Different two pulse Ramsey phase-shift:

$$\phi = \phi_{1'} + \phi_1 \qquad (8b)$$

Generalized hyper-Ramsey phase-shift:

$$\phi = \phi_{1'} + \phi_1 - \arg[\beta(\tilde{\vartheta}_{12})] \qquad (8c)$$

These analytical expressions are extending previous perturbative or exact solutions related to Ramsey and hyper-Ramsey lineshapes from refs [2,7,11]. We have reported in Fig. 2(a) the shape of the transition probability related to the original HR scheme using Eq. 4 to Eq. 6 versus the clock detuning.

Note that in order to lock the laser frequency on the atomic transition, a phase-step modulation with opposite sign $\varphi_l^\pm$ can also be applied to the ground or excited-state transition probabilities [14] generating a dispersive error signal ΔE given by:

$$\Delta E = P_g(\varphi_l^+) - P_g(\varphi_l^-) \qquad (9)$$

The subscript $l = 1', 1, 2$ is related to any kind of phase-step applied during left or right interaction pulses. Several sequences of phase-steps have been proposed with different error signal sensitivity to the composite phase-shift. The original hyper-Ramsey scheme (HR) is based on $\varphi_{1'}^\pm = \pm \pi/2$ during the first pulse with $\varphi_1 = \pi$ during the middle pulse as reported in [6,7]. Another possibility called the modified hyper-Ramsey (MHR) technique is to apply $\varphi_{1'}^+ = +\pi/2$ during the first pulse, keeping $\varphi_1 = \pi$ while using an opposite phase-step $\varphi_2^- = -\pi/2$ on the last pulse [15]. Other protocol is the generalized hyper-Ramsey (GHR) scheme where phase-steps are applied only during the middle pulse with $\varphi_1 = \pm \pi/4$ or $\varphi_1 = \pm 3\pi/4$ [16]. All error signals are easily retrieved within the spinor formalism following phase-steps protocols previously listed and are fully consistent with results already presented in ref [10].

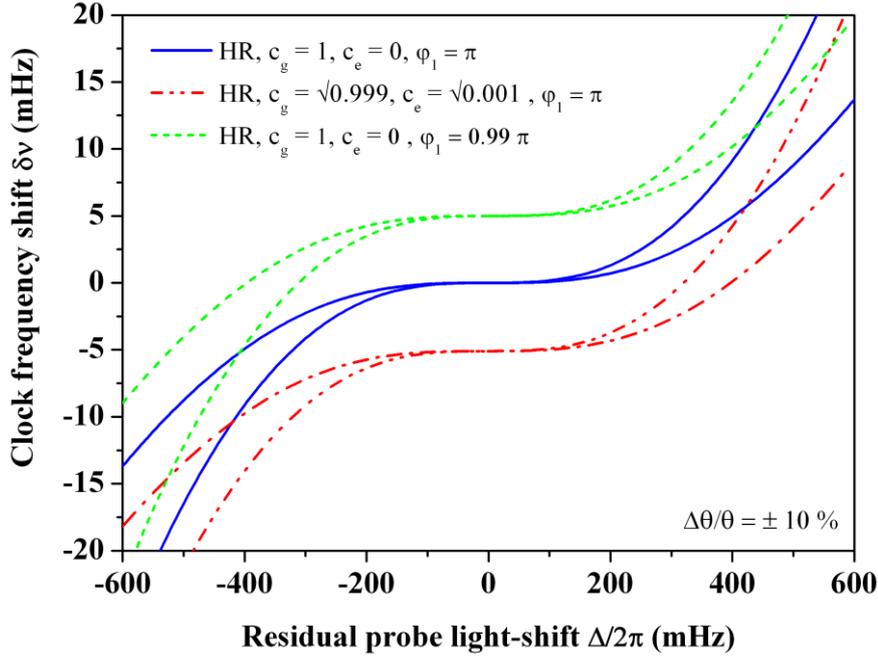

Figure 4. The HR clock frequency-shift $\delta\nu$ of the population fraction $P_e$ experiences a small offset due to unperfect state initialization and phase-step error versus residual uncompensated probe light-shift $\Delta/2\pi$. Identical colored dashed-dots and solid lines are produced by a pulse area variation of $\frac{\Delta\theta}{\theta} = \pm 10\%$. Others parameters as in Fig. 2.

## 3. GHR transition probability including arbitrary initial amplitudes

We establish the full lineshape expression of the transition probability for arbitrary initial amplitudes $c_g(0), c_e(0)$ respecting the condition $|c_g(0)|^2 + |c_e(0)|^2 = 1$. New expressions for coefficients $\alpha(\tilde{\vartheta}_{1\prime}), \beta(\tilde{\vartheta}_{1\prime})$ driving the first laser pulse excitation are now given by:

$$\alpha(\tilde{\vartheta}_{1\prime}) = \cos\tilde{\vartheta}_{1\prime}\left(c_g(0) + ic_e(0)e^{i(\varphi_{1\prime} - \phi_{1\prime})}\tan\tilde{\vartheta}_{1\prime}\right) \quad (10a)$$

$$\beta(\tilde{\vartheta}_{1\prime}) = \frac{c_g(0)\tan\tilde{\vartheta}_{1\prime} - ic_e(0)e^{i(\varphi_{1\prime} - \phi_{1\prime})}}{c_g(0) + ic_e(0)e^{i(\varphi_{1\prime} - \phi_{1\prime})}\tan\tilde{\vartheta}_{1\prime}} \quad (10b)$$

Envelop and phase-shift driving the entire GHR transition probability are replaced by complex trigonometric functions including imperfect initialization of quantum states. The generalized composite phase-shift expression becomes:

$$\phi = \phi_{1\prime} + \phi_1 - \left(\arg[\beta(\tilde{\vartheta}_{1\prime})] + \arg[\beta(\tilde{\vartheta}_{12})]\right) \quad (11)$$

where $\beta(\tilde{\vartheta}_{12})$ and $\beta(\tilde{\vartheta}_{1\prime})$ are respectively given by Eq. 6d and Eq. 10b. The clock frequency-shift associated to this phase-shift is given by the following relation:

$$\delta\nu = -\frac{\Phi + k.\pi}{2\pi\,T} \quad (12)$$

where k is a positive or negative integer selected to ensure expression continuity by correcting quadrant jumps.

We have reported in Fig. 2(b) the HR clock frequency-shift associated to the resonance minimum through Eq. 12 versus residual uncompensated probe induced shifts and pulse area in units of $\pi/2$. This is the original signature feature of the ideal HR scheme already reported in [17] with a density matrix formalism. The distortion of the HR clock frequency-shift reported in Fig. 3(a) and (b) are produced when state initialization is incorrect. Laser phase-steps are fixed to be $\varphi_{1'} = 0, \varphi_1 = \pi, \varphi_2 = 0$ for a simplified simulation. A modification of the relative phase-shift between composite pulses is due to a residual coherence between atomic states $c_g(0), c_e(0) \neq 0$ in a quantum mechanical approach. This new result might emphasize that a weak distortion of the lineshape may coexist with an additional phase-shift contribution related to a lack of appropriate quantum state engineering between multiple sequences of two-level systems probe interrogation [18]. We have finally reported in Fig. 3(c) the effect of a systematic small error in the phase-step inversion $\varphi_1 = 0.99\,\pi$ realized during the middle pulse of the HR scheme. The clock frequency shift experiences a small offset against residual probe-induced shifts if phase-steps or state initialization are not well controlled as shown in Fig. 4. The clock frequency shift is assuming a $\frac{\Delta\theta}{\theta} = \pm 10\,\%$ pulse variation during the probe interrogation.

## 4. Conclusion

We have derived the generalized hyper-Ramsey (GHR) resonance with spinors. The exact lineshape and the central fringe phase-shift have been rewritten in a compact form allowing to track any distortion from pulse defects. For the first time, a composite phase-shift is including arbitrary initial amplitude of quantum states. We have employed spinors to establish an exact and tractable model of the GHR resonance which might be very helpful when intensity fluctuations are taken into consideration [19]. Merging the original method of Ramsey spectroscopy [2] with more complex sequences of composite laser pulses [20,21] would allow optimal control of laser parameters in the field of clock interferometry and precision spectroscopy for tests of fundamental physics with ultra-cold quantum particles.

A.V. Taichenachev and V.I. Yudin acknowledge financial support from Russian Scientific Foundation (grant no. 16-12-0052) and from Russian Foundation for Basic Research (grant no. 18-02-00822)

**Names of authors:**

**Thomas Zanon-Willette**

**E-mail address: thomas.zanon@sorbonne-universite.fr**

**Sorbonne Université, Observatoire de Paris, Université PSL, CNRS, Laboratoire d'Etudes du Rayonnement et de la Matière en Astrophysique et Atmosphères, LERMA, F-75005, Paris, France**

**Alexey V. Taichenachev**

**Novosibirsk State University, ul. Pirogova 2, Novosibirsk, 630090, Russia and Institute of Laser Physics SB RAS, pr. Ak. Lavrent'eva 15 B, Novosibirsk 630090, Russia**

**Valeriy I. Yudin**

**Novosibirsk State University, ul. Pirogova 2, Novosibirsk, 630090, Russia and Institute of Laser Physics SB RAS, pr. Ak. Lavrent'eva 15 B, Novosibirsk 630090, Russia, Russia and Novosibirsk State Technical University, pr. Karla Marksa 20, Novosibirsk, 630073, Russia**